\documentclass[useAMS,usenatbib,letterpaper,onecolumn]{mn2e}
\usepackage[totalwidth=480pt,totalheight=680pt]{geometry}
\usepackage{graphicx,amssymb,amsbsy}
\newcommand{\be}{\begin{equation}}
\newcommand{\ee}{\end{equation}}
\def\lta{\,\raise 0.3 ex\hbox{$ < $}\kern -0.75 em
 \lower 0.7 ex\hbox{$\sim$}\,}
\def\gta{\,\raise 0.3 ex\hbox{$ > $}\kern -0.75 em
 \lower 0.7 ex\hbox{$\sim$}\,} 
\newcommand{\zhat}{{\hat z}} 
\newcommand{\xhat}{{\hat x}} 
\newcommand{\mtot}{m_{\scriptstyle T}} 
\newcommand{\mplan}{ m_{\rm p}} 
\newcommand{\nump}{ {\cal N} } 
\newcommand{\sound}{c_{\rm s}}  
\newcommand{\selfg}{\alpha_{\rm g}}
\newcommand{\pfact}{\alpha_{\rm p}}
\newcommand{\stepfun}{ {\cal H} }

\title[Pairwise Tidal Equilibrium in Extrasolar Planetary Systems]
{Pairwise Tidal Equilibrium States and the \\
Architecture of Extrasolar Planetary Systems} 

\author[F. C. Adams]
{Fred C. Adams \\ 
Physics Department, University of Michigan, Ann Arbor, MI 48109\\
Astronomy Department, University of Michigan, Ann Arbor, MI 48109} 

\begin{document}

\date{May 2019} 
\pagerange{\pageref{firstpage}--\pageref{lastpage}} \pubyear{2019} 
\maketitle
\label{firstpage} 

\begin{abstract} 
Current observations indicate that the planet formation process often
produces multiple planet systems with nearly circular orbits, regular
spacing, a narrow range of inclination angles, and similar planetary
masses of order $\mplan\sim10M_\oplus$. Motivated by the observational
sample, this paper determines the tidal equilibrium states for this
class of extrasolar planetary systems. We start by considering two
planet systems with fixed orbital spacing and variable mass ratios.
The basic conjecture explored in this paper is that the planet
formation process will act to distribute planetary masses in order to
achieve a minimum energy state. The resulting minimum energy
configuration --- subject to the constraint of constant angular
momentum --- corresponds to circular orbits confined to a plane, with
nearly equal planetary masses (as observed). We then generalize the
treatment to include multiple planet systems, where each adjacent pair
of planets attains its (local) tidal equilibrium state. The properties
of observed planetary systems are close to those expected from this
pairwise equilibrium configuration. In contrast, observed systems do
not reside in a global minimum energy state. Both the equilibrium
states of this paper and observed multi-planet systems, with planets
of nearly equal mass on regularly spaced orbits, have an effective
surface density of the form $\sigma\propto r^{-2}$, much steeper than
most disk models.
\end{abstract} 


\section{Introduction} 
\label{sec:intro} 

With thousands of extra-solar planets now discovered, the observed
collection of planetary systems displays an enormous degree of
diversity. On the other hand, the multiple-planet systems detected
through the {\it Kepler} mission \citep{borucki,batalha} exhibit
several instances of uniformity: The typical planetary mass in this
sample is of order 10 $M_\oplus$ \citep{zhu2018}, and planets within
the same system tend to have similar masses \citep{millholland,songhu}
and sizes \citep{weiss1,weiss2}.  The planets are found to orbit in
nearly the same plane \citep{fangmargot,tredong} and are remarkably
stable against oscillations in their orbital inclination angles (so
that they stay aligned; \citealt{beckeradams16,beckeradams17}). The
orbital eccentricities are low, much smaller than those of the
exoplanet sample taken as a whole \citep{vaneylen,hadden}. The angular
momentum vectors of the stellar spins and planetary orbits are
well-aligned in most cases \citep{winnfab}. These planetary systems
generally have regularly spaced orbits with relatively small
separations \citep{rowe2014}, but are spread out enough to remain
stable \citep{puwu}. Although the observed orbital proximity often
results in period ratios that are close to small integer values, the
systems are rarely found in bonafide low-order mean motion resonances
\citep{fabrycky}. These features of the observed planetary systems
suggests that some organizing principle is operating during the
process of planet formation, i.e., during the phase when planetary
masses and orbits are determined.  The goal of this work is to show
how considerations of energy optimization --- subject to constraints
of mass and angular momentum conservation --- provide a partial
explanation for the regularities observed in the sample of extrasolar
multi-planet systems.

A classic dynamical issue is to find the tidal equilibrium state for
astrophysical systems containing both self-gravity and angular
momentum (beginning with \citealt{darwin1,darwin2}). Given its
origins, this type of constrained optimization procedure is sometimes
known as a {\sl Darwin Problem}. Any type of dissipation causes
physical systems to evolve toward lower energy states, so that the
lowest energy (equilibrium) state is the preferred configuration.
These states are often called tidal equilibrium states because tidal
interactions cause dissipation in many astrophysical systems of
interest. For example, in binary star systems with both rotational and
spin angular momentum, the existence of an equilibrium state requires
both a minimum total angular momentum and a sufficiently large orbital
component \citep{counselman,hut1980}, and tidal interactions provide a
mechanism to reach equilibrium \citep{hut1981}.  These results for
two-body systems have been applied to Hot Jupiters \citep{levrard},
and have been generalized to include the quadrupole moment of the
central star \citep{ab2015} and a third body in hierarchical
star-planet-moon systems \citep{ab2016}.

In the aforementioned applications, the components of the system
include spin angular momentum, orbital angular momentum, and the
orbital energy; the masses of the bodies are considered fixed.
Moreover, the standard approach only allows for the inclusion of one
orbit --- that of the binary system. In the present case, however, the
systems of interest include several planets, with orbits that are
generally wide enough to be decoupled from the stellar rotation over
the lifetime of the system.  The spin angular momenta of the planets
themselves are negligible. On the other hand, one quantity of interest
is the mass ratio for adjacent pairs of planets. We thus need to solve
for the optimal distribution of planetary masses. Note that the mass
fractions will be determined through the process of planet formation,
whereas the goal of this paper is to find the preferred (minimal
energy) configuration.

This work poses a new type of Darwin problem that includes multiple
orbits and allows the masses of the bodies to vary (subject to the
constraint that the total mass $\mtot$ in planets is held constant).
All of the angular momentum is contained in the planetary orbits and
the total ${\bf L}$ is constrained to be constant. In this
formulation, the orbital spacing, denoted as $\Lambda$, is also held
fixed.  For two planet systems, the relevant variables thus include
the planet masses $m_j$, the orbital eccentricities $e_j$, the
relative inclination angle $i$ of the orbits, and the semimajor axis
$a$ of the inner planet (note that the semi-major axis of the outer
planet is set by the fixed orbital spacing $\Lambda$). The problem is
thus to optimize the system properties $(m_1,m_2,e_1,e_2,i,a)$ subject
to the constraints of constant $({\bf L},\mtot,\Lambda)$.  With the
two planet solution in hand, we then generalize to systems with more
planets. If multiple planet systems evolve toward configurations where
each adjacent pair of planets resides in a tidal equilibrium state,
the resulting systems closely resemble those that are observed (with
nearly equal mass planets and regularly spaced circular orbits in the
same plane).

The goal of this paper is to provide a partial explanation for the
properties of observed multi-planet systems by showing that they
reside near optimized energy configurations. This approach does not
describe the evolutionary path by which they attain such states.
Instead, this calculation highlights a key physical principle ---
namely energy minimization --- that is likely to constrain any
scenario for planetary assembly.  As a result, this work is
complementary to previous studies of planet population synthesis
(e.g., \citealt{mordasini,idalin,alessi,mordasini18}) that model the
sequence of evolutionary steps, including rock agglomeration, core
formation, gas accretion, planetary migration, and so on.

This paper is organized as follows. We first consider the case of a
two planet system in Section \ref{sec:double}.  We find the resulting
tidal equilibrium state and show that it is a minimum. This treatment
is then generalized in Section \ref{sec:multiple} to include systems
with multiple planets. We show that adjacent pairs of planets can
attain minimum energy states, and that the resulting system
architecture of the equilibrium configuration provides a good
description of observed planetary systems.  
{Using a sample of observed multi-planet systems, Section
\ref{sec:observe} considers how well the energy scales found in
observed pairs conform to the theoretical expectations of pair-wise
equilibrium states.}  The paper concludes in Section
\ref{sec:conclude} with a summary of results and a discussion of their
implications. We also show that no global energy minimum exists if the
total planetary mass can be distributed among three or more bodies
(Appendix \ref{sec:equilthree}). Finally, we present an order of
magnitude estimate for the expected planetary masses based on the
concept of pebble accretion (Appendix \ref{sec:rockmass}).

\section{Two Planet Systems} 
\label{sec:double} 

Here we consider a system containing two planets. The goal is to 
find the lowest energy state of the two planet system, subject to
conservation of planetary mass and total system angular momentum.  
The mass fractions of the planets are allowed to vary: The idea is
to find the state where the planets optimize energy during their 
formation process. 

The planetary spins are assumed to have negligible angular momentum,
so that the total angular momentum of the system is constant and is
determined by the orbital angular momentum of the two planets. Without
loss of generality, we can take the angular momentum vector of the
inner planet to point in the $\zhat$ direction, so that the total
angular momentum can be written in the form 
\be
{\bf L} = h_1 \zhat + h_2 
\left[ \zhat \cos i + \xhat \sin i \right] \,, 
\ee
where $h_j$ is the orbital angular momentum of each planet. The
inclination angle $i$ is measured from the pole. Here we take the
magnitude of the angular momenta to be given by the expression in the
low mass limit, i.e., 
\be
h_j^2 = m_j^2 G M_\ast a_j (1 - e_j^2) \,, 
\ee
where $M_\ast$ is the stellar mass, $m_j$ is the planetary mass, 
and $e_j$ is the orbital eccentricity (i.e., we assume that 
$m_j\ll{M_\ast}$). The energy of the system is given by 
\be
E = - {G M_\ast m_1 \over 2a_1} - {G M_\ast m_2 \over 2a_2} \,. 
\ee
Note that, in general, the orbits for each of the two planets are
described by six orbital elements. This treatment considers only 
the energy and angular momentum of the orbits. The interactions 
between the planets are assumed to determine the orbital spacing 
(through the parameter $\Lambda$), but are not included in the 
energy budget for the optimization procedure. This correction to 
the energy is of order $m_j/[M_\ast(\Lambda-1)]\sim 10^{-4}$.  

Here we invoke a constraint on the total mass of the planetary 
pair, such that 
\be
m_1 + m_2 = \mtot = constant\,, 
\ee 
but allow the individual masses to vary. Motivated by the tendency for
planetary migration to push planetary orbits together, we enforce a
second constraint on the orbital spacing.  We thus consider systems 
with a fixed separation, corresponding to a given number $K$ of Hill
radii $R_H$, so that 
\be
a_2 = a_1 (1 + \Delta) = a_1 \Lambda \,,
\ee
where 
\be
\Delta = K R_H/a_1 \qquad {\rm and} \qquad 
R_H = \left( {\mtot \over 3M_\ast} \right)^{1/3} a_1 \,.
\ee
The minimum value of $K$ required for stability in two-planet systems
is generally taken to be $K=2\sqrt{3}$ \citep{gladman}, whereas larger
values are required for systems with more planets. Numerical
simulations suggest that $K\sim10$ or larger is needed for long-term
stability \citep{chambers}, where the value depends on the number and
masses of the planets \citep{obertas,wu2019}. For comparison, the
observed separations of planetary orbits in the {\it Kepler} sample
display a wide distribution with a broad peak in the range $K=10-20$
(e.g., compare \citealt{weiss1} and \citealt{puwu}). Accounting for
variations in the masses of the host stars, this range in $K$
corresponds to spacing parameters in the approximate range
$\Lambda\approx1.2-1.8$ (see also Section \ref{sec:spacing} and 
Figure \ref{fig:xlhist} below).

Next we define dimensionless quantities according to  
\be
f = {m_1 \over \mtot}, \qquad 1-f = {m_2\over \mtot}, \qquad
a = {a_1 \over R_\ast}, \qquad a \Lambda = {a_2 \over R_\ast} \,. 
\ee
The energy can be written in terms of the energy scale 
$GM_\ast\mtot/2R_\ast$ and the angular momentum can be written 
in terms of the scale $\mtot (G M_\ast R_\ast)^{1/2}$. After 
dividing out these scales, the dimensionless expressions for 
the energy and angular momentum become 
\be 
E = - {1\over a} \left[ f + {1-f \over \Lambda} \right]
\ee
and 
\be
{\bf L} = \sqrt{a} \left\{ f (1-e_1^2)^{1/2} \zhat + 
(1-f) \sqrt{\Lambda} (1-e_2^2)^{1/2} \left[ \zhat \cos i + 
\xhat \sin i \right] \right\} \,. 
\ee

\subsection{Extremum of the Energy} 

For fixed orbital spacing (given by $\Lambda$), the system defined
above is described by five variables: the semimajor axis $a$ of the
inner planet, the mass fraction $f$ of the inner planet, the orbital
eccentricities $e_j$, and the inclination angle $i$. The system is
also subject to conservation of angular momentum. The problem is thus
to find the values of the variables $(f,a,e_1,e_2,i)$ that minimize
the energy $E$ subject to the constraint ${\bf L}$ = {\sl constant}.

The first step is to to extremize the energy $E$ subject to the
constraint that the angular momentum is constant using Lagrange
multipliers $\lambda_z$ and $\lambda_x$ (one for each component of
the angular momentum). For each variable $\xi_k$, we must enforce 
the optimization condition 
\be
{\partial E \over \partial \xi_k} + 
\lambda_z {\partial L_z \over \partial \xi_k} + 
\lambda_x {\partial L_x \over \partial \xi_k} = 0 \,, 
\ee
where the $\xi_k$ represent the variables $(f,a,e_1,e_2,i)$. We 
thus obtain five equations that specify the equilibrium state:

\noindent
[1] For the semimajor axis $a$ we obtain  
\be
{1 \over a^2} \left[ f + {1-f \over \Lambda} \right] + 
{\lambda_z \over 2 a^{1/2}} 
\left[ f(1-e_1^2)^{1/2} + 
(1-f)\sqrt{\Lambda}(1-e_2^2)^{1/2}\cos i \right]  
+ {\lambda_x \over 2 a^{1/2}} 
(1-f)\sqrt{\Lambda}(1-e_2^2)^{1/2}\sin i = 0 \,. 
\label{opta} 
\ee 

\noindent 
[2] For the mass fraction $f$ we obtain 
\be
- {1 \over a} \left[ 1 - {1\over\Lambda} \right] 
+ \lambda_z \sqrt{a} 
\left[ (1-e_1^2)^{1/2} - \sqrt{\Lambda} (1-e_2^2)^{1/2} \cos i \right] 
- \lambda_x \sqrt{a} \sqrt{\Lambda} (1-e_2^2)^{1/2} \sin i = 0\,. 
\label{optf} 
\ee

\noindent 
[3] For the orbital eccentricity $e_1$ of the inner 
planet, the optimization condition takes the form 
\be
- \lambda_z \sqrt{a} f {e_1 \over (1-e_1^2)^{1/2} } = 0 \,.
\ee

\noindent 
[4] Similarly, for the orbital eccentricity $e_2$ of the outer 
planet, we obtain 
\be
- \sqrt{a} (1-f) \sqrt{\Lambda} 
\left[ \lambda_z \cos i + \lambda_x \sin i \right] 
{e_2 \over (1-e_2^2)^{1/2} } = 0 \,. 
\ee

\noindent
[5] Finally, for the inclination angle $i$, we find
\be
\sqrt{a} (1-f) \sqrt{\Lambda} (1-e_2^2)^{1/2} 
\left[ -\lambda_z \sin i + \lambda_x \cos i \right] = 0 \,. 
\ee

The optimization conditions can be solved to find $e_1=0=e_2$, $i=0$,
and $\lambda_x=0$. The optimal solution thus corresponds to circular
orbits in the same orbital plane. The remaining equations (\ref{opta})
and (\ref{optf}) can both be solved for $\lambda_z$: 
\be 
- \lambda_z a^{3/2} = 
2 \left[ f + {1-f \over \Lambda} \right] 
\left[ f + (1-f) \sqrt{\Lambda} \right]^{-1}\,,
\ee 
and 
\be
- \lambda_z a^{3/2} = 
\left[ 1 - {1\over \Lambda} \right]
\left[ \sqrt{\Lambda} - 1 \right]^{-1}\,. 
\ee
Combining these two expressions, we can solve for the mass 
fraction $f$ to obtain 
\be 
f = {\Lambda + \sqrt{\Lambda} - 2 \over 3 (\Lambda-1)} \,.
\label{mfract} 
\ee
In the limit $\Lambda\to1$ ($\Delta\to0$), we obtain the expansion 
\be
f \to {1 \over 2} - {\Delta \over 24} + {\Delta^2 \over 48} + \dots\,,
\ee
which corresponds to the mass being (nearly) equally distributed
between the two planets. Moreover, even for relatively large values of
the separation parameter $\Delta$, the fraction $f$ remains close to
1/2. In the opposite limit $\Lambda\to\infty$, the mass fraction
$f\to1/3$. For completeness, we can also find the ratio of planetary 
masses: 
\be
{m_1 \over m_2} = {f \over 1-f} = 
{\sqrt{\Lambda} + 2 \over 2 \sqrt{\Lambda} + 1} \,. 
\label{mratio} 
\ee

Although the main result of this analysis is that the planetary masses
must be nearly equal when the system resides at its critical point, to
higher order we find that the inner planet is somewhat smaller than
the outer planet.  This state of affairs is seen in the sample of
extrasolar planetary systems: The radii of observed planets within a
given system tend to increase (slowly) with their orbital separation
\citep{kipping}, e.g., the outer planet is larger for $\sim65\%$ of 
the observed cases \citep{weiss1}. 

Given the solution (\ref{mfract}) for the mass fraction, and the 
optimization conditions ($e_1=e_2=i=0$), the magnitude $L$ of the 
angular momentum is related to the semimajor axis according to
\be
L = \sqrt{a} {2 \over 3} {\Lambda + \sqrt{\Lambda} + 1 
\over \sqrt{\Lambda} + 1 } \,. 
\ee
The lowest energy state can then be written in the form 
\be
E = - {4 \over 9L^2} \left[{\Lambda + \sqrt{\Lambda} + 1 
\over \sqrt{\Lambda} + 1 } \right]^2 
\left[ {\Lambda + \sqrt{\Lambda} + 1 \over 3\Lambda} \right] \,.  
\ee
Note that in the limit of large $\Lambda\gg1$, this expression for the
energy has the asymptotic form $E\propto-\Lambda$. In other words, the
system can always evolve to a lower energy state by spreading out the
planets, i.e., moving one planet inward and the other outward in such
a way as to conserve angular momentum. 

In the limits where either $f\to0$ or $f\to1$, the energy of the
system becomes $E=-1/L^2$. The depth of the energy minimum for 
the equilibrium state (where $f$ is given by equation [\ref{mfract}]), 
can thus be measured by the quantity 
\be
{\Delta{E}\over E} = {4 (\Lambda + \sqrt{\Lambda} + 1)^3 
\over 27 \Lambda (\sqrt{\Lambda} + 1)^2} - 1 = 
{(\sqrt{\Lambda}-1)^2 (2+\sqrt{\Lambda})^2 (1+2\sqrt{\Lambda})^2 
\over 27 \Lambda (\sqrt{\Lambda} + 1)^2} \to 
{3 \Delta^2 \over 16} + \dots \,. 
\label{pairenergy} 
\ee
The final expression shows the result in the limit where 
the spacing factor $\Lambda$ is close to unity so that 
$\Delta\ll1$. 

\subsection{Second Variation} 

The above analysis defines the critical point of the system. To show
that the critical point corresponds to a minimum of the energy, we
must consider the second variation. As is well known, the critical
point will be a minimum if and only if the eigenvalues of the Hessian
matrix $\mathbb{H}$ are all positive \citep{hesse}, where the matrix
elements are given by
\be
\mathbb{H}_{jk} = 
{\partial^2 E \over \partial \xi_j \partial \xi_k} \,. 
\label{hmatrix} 
\ee
This section shows that the eigenvalues are indeed positive, so that
the critical point represents the energy minimum.

To evaluate the Hessian matrix, we explicitly incorporate 
conservation of angular momentum into the expression for the
energy. For convenience, we define reduced dimensionless 
angular momenta for the orbits:
\be
\ell_1 \equiv f (1-e_1^2)^{1/2} \qquad {\rm and} \qquad 
\ell_2 \equiv (1-f) \sqrt{\Lambda} (1-e_2^2)^{1/2} \,.
\ee
The magnitude of the total angular momentum is then given by 
\be
L^2 = a \left[ \ell_1^2 + \ell_2^2 + 2 \ell_1 \ell_2 \cos i \right]\,,
\ee
so that the energy takes the form 
\be
E = -{1\over L^2} 
\left[ \ell_1^2 + \ell_2^2 + 2 \ell_1 \ell_2 \cos i \right]
\left[ f + {1-f \over \Lambda} \right] \,. 
\label{enercons} 
\ee
This formulation of the problem explicitly conserves angular 
momentum and eliminates the semimajor axis $a$ (of the inner
planet). The energy is thus a function of four variables 
$E=E(f,e_1,e_2,i)$. 

We now re-calculate the derivatives and the second derivatives.  In
this case, all of the mixed derivatives are zero when evaluated at the
critical point, so that the eigenvalues are given by the diagonal
terms of the Hessian matrix. Notice also that the factor of $L^2$ in
equation (\ref{enercons}) represents a multiplicative constant, and
can be ignored for purposes of testing for stability (we thus set
$L=1$ for the following analysis).

The derivative of the energy with respect to the mass fraction 
$f$ takes the form 
\be
{\partial E \over \partial f} =  
- 2 \left[ (\ell_1 + \ell_2 \cos i) (1-e_1^2)^{1/2} - 
(\ell_2 + \ell_1 \cos i) \sqrt{\Lambda} (1-e_2^2)^{1/2}\right] 
\left[ f + {1-f \over \Lambda} \right] 
- \left[ \ell_1^2 + \ell_2^2 + 2 \ell_1 \ell_2 \cos i \right]
\left[ 1 - {1\over \Lambda} \right] \,. 
\ee
Solving this equation at the critical point where
$\partial{E}/\partial{f}$ = 0 (and where $e_j=0=i$ from before), 
we obtain the mass fraction of equation (\ref{mfract}), as expected.
The second derivative, evaluated at the critical point, can be written
\be
{\partial^2 E \over \partial f^2}\Bigg|_0 = 
-2 (1 - \sqrt{\Lambda})^2 \left[ f + {1-f \over \Lambda} \right] 
- 4 (1 - \sqrt{\Lambda}) \left[ f + (1-f) \sqrt{\Lambda} \right] 
\left[ 1 - {1\over \Lambda} \right] \,. 
\ee 
Using the solution for the mass fraction $f$ at the critical point, 
this expression simplifies to the form 
\be
{\partial^2 E \over \partial f^2}\Bigg|_0 = 2 (\sqrt{\Lambda}-1)^2 
{\Lambda + \sqrt{\Lambda} + 1 \over \Lambda} \,,
\label{energyff} 
\ee 
which is manifestly positive. 

Similarly, for the inclination angle 
the first derivative has the form 
\be
{\partial E \over \partial i} = 2 \ell_1 \ell_2 \sin i 
\left[ f + {1-f \over \Lambda} \right] = 0 \,. 
\ee
We thus obtain $i=0$ as before. The second derivative becomes
\be
{\partial^2 E \over \partial i^2}\Bigg|_0 = 2 f (1-f) \sqrt{\Lambda} 
\left[ f + {1-f \over \Lambda} \right] \,. 
\label{energyii} 
\ee
This quantity if positive for all values of $f$, including the 
value at the critical point given by equation (\ref{mfract}).  

The derivatives for the eccentricities take the form 
\be
{\partial E \over \partial e_1} = \left[ f + {1-f\over\Lambda}\right] 
\left[ 2 \ell_1 + 2 \ell_2 \cos i \right] f {e_1 \over (1-e_1^2)^{1/2}} 
= 0 \,, 
\ee
and 
\be
{\partial E \over \partial e_2} = \left[ f + {1-f\over\Lambda}\right] 
\left[ 2\ell_2 + 2 \ell_1 \cos i \right] (1-f) \sqrt{\Lambda} 
{e_2 \over (1-e_2^2)^{1/2}} = 0 \,. 
\ee
The derivatives vanish for $e_1=0$ and $e_2=0$. The corresponding 
second derivatives become 
\be
{\partial^2 E \over \partial e_1^2}\Bigg|_0 = 
2 \left[ f + {1-f\over\Lambda}\right] 
\left[ f + (1-f) \sqrt{\Lambda} \right] f \,,
\label{energyee1} 
\ee
and 
\be
{\partial^2 E \over \partial e_2^2} \Bigg|_0 = 
2 \left[ f + {1-f\over\Lambda}\right] 
\left[ f + (1-f) \sqrt{\Lambda} \right] (1-f) \sqrt{\Lambda} \,, 
\label{energyee2} 
\ee
both of which are positive.

In summary, we find that the Hessian matrix is diagonal, with all four
eigenvalues manifestly positive. As a result, the critical point is a
minimum of the energy. In this minimum energy configuration, the
planets have circular orbits ($e_j=0$) in the same plane ($i=0$), with
a mass ratio close to unity. More specifically, the mass fraction $f$
of the inner planet varies slowly with the orbital separation and is
given by equation (\ref{mfract}).

\begin{figure} 
\centerline{ 
\includegraphics[width=0.70\textwidth]{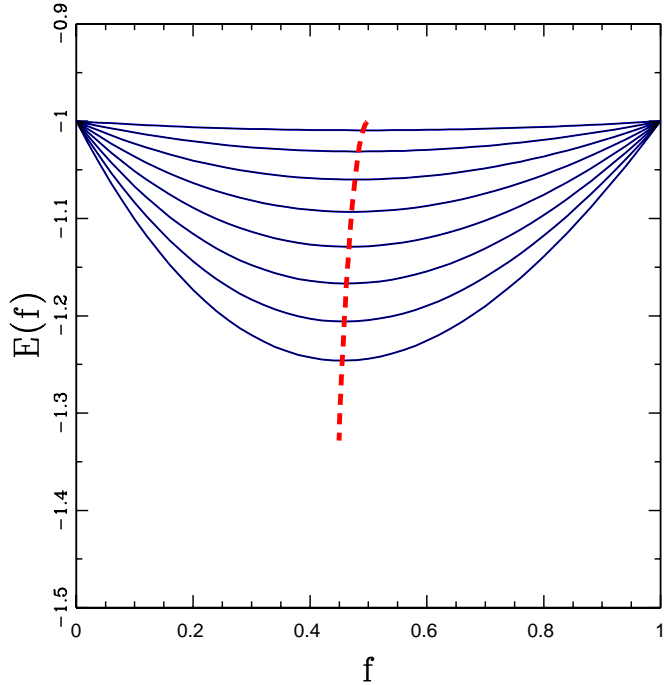} } 
\vskip-1.2truein
\caption{System energy as a function of mass fraction $f$ for a
collection of orbital spacings. The blue curves show the energy in
dimensionless units --- scaled to the orbital energy for the limiting
case where all the mass resides in a single planet. Curves are given
for $\Lambda$ = 1.25 -- 3.0 (in increments of 0.25, from top to bottom
in the diagram). The red dashed curve shows the locus of energy minima
as a function of mass fraction. The minimum slowly shifts to lower
values of $f$ as the spacing parameter $\Lambda$ increases, but
remains close to $f\sim1/2$. }
\label{fig:energy} 
\end{figure} 

The existence and depth of the energy minimum is illustrated in Figure
\ref{fig:energy}, which shows the energy for a two planet system as a
function of the mass fraction $f$ for the inner planet.  In this case,
the eccentricities and inclination angle are set to zero, the values
appropriate for the equilibrium state.  With the semimajor axis
determined by the requirement of conservation of angular momentum, the
energy is a function of the mass fraction $f$ only for fixed orbital
spacing $\Lambda$. The collection of curves in the figure shows the
dimensionless energy versus $f$ for orbital spacings in the range
$\Lambda$ = 1.25, 1.5, 1.75, 2.0, 2.25, 2.5, 2.75, and 3.0 (from top
to bottom).  Each curve has a minimum corresponding to the lowest
energy state. The locus of minima is marked by the red dotted curve,
where the value of mass fraction $f$ for the minimum is given by
equation (\ref{mfract}).

\section{Systems with Three or More Planets} 
\label{sec:multiple} 

The next step is to consider systems with more than two planets.
However, the optimization procedure used for the two planet model
cannot be immediately generalized for three planets. If we allow for
the total mass to be distributed among three bodies, then no solution
for the energy minimum exists. This result is shown in Appendix
\ref{sec:equilthree}. The equilibrium state requires two of the planets
to occupy the same orbit, thereby reducing the system to having only
two planets. This argument can be generalized to systems with more
than three planets. As a result, multiple systems with $N\ge3$ planets
cannot reside in a configuration that represents a global minimum of
energy (subject to constant angular momentum). Some additional
constraint must be placed on the system. This section introduces the
concept of {\sl pairwise equilibrium}, where each pair of planets on
adjacent orbits resides in the tidal equilibrium state for the two
planet problem outlined in the the previous section.

We start with three planet systems, where the three bodies are
considered to be the sum of two sets of two planets, where the middle
planet belongs to both the inner pair and the outer pair. If the
energy is minimized subject to conservation of angular momentum of
both the inner pair and (separately) the outer pair, then a solution
can be found. We denote this state of the system as pairwise
equilibrium. For both planetary pairs, the orbital spacing is assumed
to be fixed, i.e., determined by physics independent of the
optimization procedure.

Note that each of the two pairs of planets has a fixed orbital spacing
for purposes of determining the (individual) equilibrium states, but
two possibilities exist: In the first case, the ratio $\Lambda$ of the
semimajor axes is the same for both pairs.  In the second case, the
number $K$ of mutual Hill radii is the same. In this second case, the
ratio $\Lambda$ could be different for the two planetary pairs because
the masses are not necessarily the same ($m_1+m_2\ne m_2+m_3$). 
However, the difference between the two scenarios leads to corrections
that are higher order in the parameter $\Delta$, so we consider the
case of constant $\Lambda$ for simplicity.

In this case, for fixed orbital spacing $\Lambda$, the solution for
the mass fraction $f$ will be the same for both pairs, and can be
written in the form given by equation (\ref{mfract}).  For the
stationary state, each planet has an orbital angular momentum $L_k$. 
The ratio of the angular momenta of the two inner planets is given by
\be
{L_2 \over L_1} = \sqrt{\Lambda}\,\,{1 - f \over f} = 
\sqrt{\Lambda} \,\, {2\Lambda - \sqrt{\Lambda} - 1 \over
\Lambda + \sqrt{\Lambda} - 2} = \sqrt{\Lambda} \,\, 
{2\sqrt{\Lambda}+1 \over \sqrt{\Lambda} + 2} \equiv 
F(\Lambda) \,. 
\label{fdef} 
\ee
Similarly, the second pair of planets must obey the relation 
\be
{L_3 \over L_2} = F(\Lambda) \,.
\ee
The total angular momentum is thus given by 
\be
L_T = L_1 + L_2 + L_3 = L_1 (1 + F + F^2) \,. 
\label{totalone} 
\ee
Given the form of the angular momentum expression, 
consistency also requires that 
\be
L_2 + L_3 = G (L_1 + L_2) \qquad {\rm where} \qquad 
G = \sqrt{\Lambda}\,\,{m_2 + m_3 \over m_1 + m_2} \,. 
\label{gdef} 
\ee
This expression keeps track of the fact that the inner orbit for the
outer problem must be the outer orbit of the inner problem (hence the
factor of $\sqrt{\Lambda}$), and allows for the masses of the inner
planet pair and the outer planet pair to be different.  Using this
second expression, we can also write the total angular momentum in the
form
\be
L_T = L_1 + G (L_1 + L_2) = L_1 (1 + G + GF) \,. 
\label{totaltwo} 
\ee
Equating the two expressions (\ref{totalone}) and (\ref{totaltwo}) 
for the total angular momentum, we can derive a relationship 
between the functions $F$ and $G$, i.e., 
\be
G + GF = F + F^2 \qquad {\rm or} \qquad G=F\,.
\ee
For completeness, note that the above equation is quadratic in $F$ and
has two roots. The second solution $F=-1$ allows for equality for any
value of $G$, but is unphysical.

For a given spacing of orbits $\Lambda$, the mass ratio of 
the inner and outer pairs has the form 
\be
{m_2 + m_3 \over m_1 + m_2} = 
{2\sqrt{\Lambda}+1 \over \sqrt{\Lambda} + 2} \equiv \eta\,, 
\label{etadef} 
\ee
where we have defined the parameter $\eta$ in the final 
equality. After some algebra, the mass fractions for the 
three planets can be written in the form 
\be
{m_1 \over m_T} = {1 \over 1 + \eta + \eta^2}\,, \qquad 
{m_2 \over m_T} = {\eta \over 1 + \eta + \eta^2}\,, \qquad 
{\rm and} \qquad 
{m_3 \over m_T} = {\eta^2 \over 1 + \eta + \eta^2}\,.
\label{etamass} 
\ee
We can gain some insight by writing these expressions to leading 
order in $\Delta$, where the mass fractions become 
\be
{m_1 \over m_T} = {1\over3} \left(1-{\Delta\over6}\right)\,, \qquad 
{m_2 \over m_T} = {1\over3}\,, \qquad {\rm and} \qquad 
{m_3 \over m_T} = {1\over3} \left(1+{\Delta\over6}\right)\,.
\ee
Since $\Delta$ is small in practice (typically $\Delta\sim1/2$, 
as discussed below), the mass ratios are close to unity. 

This argument can be readily generalized for more than three
planets. For constant orbital spacing $\Lambda$, each successive
angular momentum is larger than the previous one by a factor of
$F(\Lambda)$ from equation (\ref{fdef}), so that
$L_{n+1}=F(\Lambda)L_n$. Each successive pair of angular momenta will
be larger than the previous pair by a factor of $G$ from equation
(\ref{gdef}). Provided that the system with $N$ planets has $G=F$ (as
shown above for $N=3$), then it can be shown through a straightforward
induction argument that the final pair in a system of $N+1$ planets
will also have $G=F$. As a result, in a state of pairwise
equilibrium, each planet is larger than its inner neighbor by a factor
of $\eta$ given by equation (\ref{etadef}). The mass fractions for 
the planets in an $n$ planet system can thus be written in the form 
\be
{m_k \over \mtot} = {\eta^{k-1} \over {\cal M}} \qquad 
{\rm where} \qquad {\cal M} \equiv \sum_{j=1}^{N} \eta^{j-1} \,.
\ee

\section{Comparison to Observed Planetary Systems} 
\label{sec:observe} 

This section uses the properties of the observed multiple planet
systems, primarily discovered via the {\it Kepler} mission, to 
compare with the theoretical expectations for systems in pairwise
equilibrium. After identifying the observed planetary sample of
interest, we construct the equivalent surface density distribution for
multiple-planet systems (Section \ref{sec:sigma}). The resulting mass
profile differs from the minimum mass solar nebular (MMSN), but is 
{consistent} with that predicted from pairwise equilibrium.  
We then assess the energy budget for planetary systems (Section
\ref{sec:obsenergy}) and show that the energy minimized through
pairwise equilibrium states is generally larger than the self-gravity
of the planets and the interaction energy of the planet-planet
potential. Finally, we assess the distribution of orbital spacings 
found in observed planetary systems and derive an order-of-magnitude 
estimate for the expected range (Section \ref{sec:spacing}). 

A number of previous studies have made detailed statistical analyses
of the {\it Kepler} multi-planet systems of interest (e.g., see
\citealt{millholland,weiss1,weiss2}, and references therein). For
comparison with the theoretical expectations derived in the previous
sections, we use a simplified version of this observational sample,
which is constructed as follows: From the publicly available
database\footnote{https://exoplanetarchive.ipac.caltech.edu} (where
the data are taken from a variety of sources, e.g., \citealt{batalha}), 
we consider the 219 planetary systems with 3 or more planets.  This
data set thus includes a total of 777 planets.  Most of these planets
are discovered through transit observations, some are detected through
radial velocity measurements, and some have both types of data. For
this simplified treatment, the planets with measured radii but without
reported masses are assigned masses through the empirical relation
$\mplan \sim R_{\rm p}^{2}$. Similarly, for the planets with measured
masses but no reported radii, we use the inverse of this relation to
specify $R_{\rm p}$. Although this approximate relation is sufficient
for the rough comparison used here, one should keep in mind that
planets with a given radius can have a range of masses (e.g.,
\citealt{wolfgang}).  For the stars that do not have masses listed, we
use the main-sequence relationship between stellar mass and effective
temperature to assign a mass (which is required to estimate the
semimajor axis from the measured period). We thus obtain the set of
variables $(\mplan,R_{\rm p},a,P_{\rm orb},M_\ast)$ for each planet.

Although the observational sample includes all of the multi-planet
systems that are currently detected (with $N_{\rm p}\ge3$), the data
are both heterogeneous and subject to observational bias. The sample
includes planetary systems orbiting stars with a range of masses.
However, it is likely that planet formation proceeds differently
around solar-type stars and low-mass stars (e.g., \citealt{lba}).
Unfortunately, with only $\sim200$ systems, not enough data exist to
construct separate samples for different stellar masses. The current
collection includes planets detected by both transit observations and
radial velocity measurements, and the estimates for planetary masses
are subject to well-known uncertainties. In addition, planets are more
readily detected in close orbits, so that the sample is likely to be
missing planets with longer periods. In the present application,
however, we are primarily interested in the properties of planetary
pairs within the same system. These pairs have the same stellar host
and are almost always detected by the same method, so that the
aforementioned complications are partially mitigated.

\subsection{Surface Density for Observed Multiple Planet Systems} 
\label{sec:sigma} 

\begin{figure} 
\centerline{ 
\includegraphics[width=0.70\textwidth]{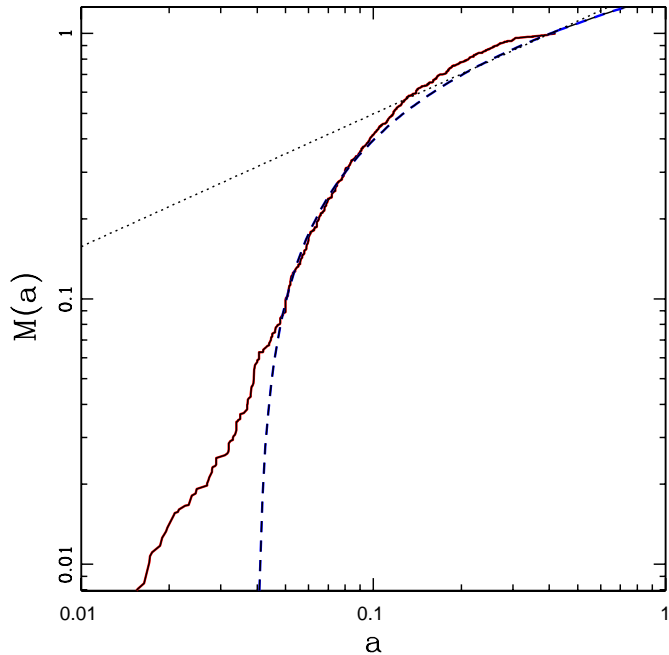} } 
\vskip-1.2truein
\caption{Mass profile of observed multiple planet systems compared
with models. The red curve shows the cumulative mass distribution for
the rocky component of the planets found in multiple planet systems
using the observational sample. This profile is limited to planets
with periods $P\le100$ days. For planets with masses
$\mplan>15M_\oplus$, only 15 $M_\oplus$ of the mass is assumed to be
composed of rock. The blue curve shows the mass distribution
$M(a)\propto\ln(a)$ appropriate for planets with uniform spacing and
equal masses, as predicted by the condition of pairwise
equilibrium. The dotted line shows the mass distribution
$M(a)\propto{a}^{1/2}$ appropriate for the benchmark surface density
profile $\sigma\propto{a}^{-3/2}$, corresponding to an extended
Minimum Mass Solar Nebula and the proposed Minimum Mass Extrasolar
Nebula. }
\label{fig:massprofile} 
\end{figure}

If planetary systems achieve or approach the minimum energy states
derived in the previous sections, then their corresponding surface
density distributions must have a particular form. {Here the
surface density is that produced by smearing out the planetary masses
into a smooth profile (e.g., see \citealt{chiang}). This subsection
constructs the mass profile and surface density distribution predicted
by pairwise equilibrium states, which have nearly equal mass planets
and regularly spaced orbits. Such a configuration has a surface
density profile $\sigma\sim a^{-2}$, as shown below. We also include a
comparison to the surface density profile constructed from observed
multi-planet systems (see Figure \ref{fig:massprofile}). }

For a given orbital spacing $\Lambda$, the planetary masses for a
pairwise equilibrium state are nearly equal. As a result, at the
location in the disk corresponding to the $N$th planet, the semimajor
axis and enclosed mass are given by 
\be 
a_N = \Lambda^N a_1 \qquad {\rm and} \qquad 
M(a_N) = N m_1 \,,
\ee
where $a_1$ and $m_1$ are the semimajor axis and mass of the first
(innermost) planet. For purposes of finding a benchmark mass distribution
for comparison, we consider $N$ to be a continuous variable, so that 
\be
N = {\ln(a/a_1) \over \ln\Lambda} \qquad {\rm and} \qquad 
M(a) = {m_1 \over \ln\Lambda} \ln(a/a_1) \, . 
\label{masslog} 
\ee
The surface density corresponding to this mass distribution can 
be found from the defining relation 
\be
{dM \over da} = 2\pi{a} \sigma(a) \qquad \Rightarrow \qquad 
\sigma(a) = \left({m_1 \over 2\pi\ln\Lambda}\right){1\over a^2} 
\propto a^{-2}\,. 
\label{sigmar2} 
\ee
This surface density distribution is thus significantly steeper 
than the usual MMSN where $\sigma \sim r^{-3/2}$ (compare with
\citealt{weiden,hayashi,desch}) and steeper than the result
$\sigma\sim r^{-1}$ often inferred for circumstellar disks associated
with newly formed stars (e.g., \citealt{andrews,perez}). This profile
is also steeper than that of the minimum mass extrasolar nebula
\citep{chiang}, which was constructed from the entire sample of
exoplanets, not just those found in multiple planet systems. 

It is important to keep in mind that equation (\ref{sigmar2})
corresponds to the effective surface density distribution for the
rocky component of planets at their current locations. If planetary
systems evolve toward lower energy states during the process of planet
formation by distributing the available mass in an optimal manner (see
Sections \ref{sec:double} and \ref{sec:multiple}), then they will
attain nearly equal masses and the final, cumulative mass
distribution of equation (\ref{masslog}).

Figure \ref{fig:massprofile} compares the mass profile expected from a
pairwise equilibrium state with that constructed from the observed
bodies in multiple planet systems. The blue dashed curve shows the
mass distribution $M(a)\sim\ln(a)$ from equation (\ref{sigmar2}),
which results from equal mass planets with uniform spacing (where we
specify the inner radius $a_1$ = 0.04 AU).  The red curves shows the
cumulative mass distribution for the observed sample. This
observational profile was constructed as follows: The planets in
the sample were first ordered according to their semimajor axes. Since
we are interested in the mass profile of the rocky component, planets
with masses larger than 15 $M_\oplus$ were assumed to have only 15
$M_\oplus$ of rocky material. The mass profile is then given by the
straightforward expression
\be
M(a) = \sum_{k=1}^{N_T} m_k \stepfun (a-a_k) \,,
\ee
where $(a_k,m_k)$ are the masses and semimajor axes of the observed
planets (subject to $m_k\le15M_\oplus$) and $\stepfun(x)$ is the
Heaviside step function \citep{abrasteg}. {Only planets with 
periods less than 100 days are included.} Both the theoretical and 
observed mass distributions are then normalized to unity at the
outermost value of $a$. Notice that the observed distribution 
{is roughly consistent with} that expected from pairwise 
equilibrium. One interesting point of departure occurs for small radii
$a\lta0.04$ AU, where the observational profile shows a weak power-law
tail; this portion of the distribution contains an excess of $\sim7\%$
of the total mass.  Both the observed mass distribution and that
predicted by pairwise equilibrium are clearly distinct from the
power-law form of the MMSN.  This latter profile is depicted in Figure
\ref{fig:massprofile} by the dotted line, which has the form
$M(a)\sim{a^{1/2}}$, corresponding to the surface density distribution
$\sigma(a)\sim{a^{-3/2}}$.

The similarity between the observed mass profile and the theoretical
construction of equation (\ref{sigmar2}) is not unexpected: This
agreement means that the observed sample of multi-planet systems
displays nearly equal mass planets with regularly spaced orbits as
claimed previously (e.g., \citealt{rowe2014,millholland,weiss1,weiss2}). 
If we enforce regular orbital spacing, then the condition of pairwise
equilibrium requires nearly equal planetary masses, as observed.

Nonetheless, the observational profile of Figure \ref{fig:massprofile}
is subject to a number of uncertainties: Given the limited number of
planets detected in multi-planet systems (777 in this sample), we have
included planets orbiting host stars of all masses. The surface
density profiles could display some dependence on stellar mass, and
that trend is not captured by this compilation. This construction also
implicitly assumes that the planetary systems are complete out to
periods $P=100$ days. Additional planets, as yet undetected, could
also be present in these systems, and would alter the inferred mass
distribution. More specifically, biases in planet detection favor the
discovery of short-period planets. The apparent absence of additional
long-period planets could contribute to the inferred steepness of the
surface density distribution found here. 

\subsection{Energy Scales} 
\label{sec:obsenergy} 

\begin{figure} 
\centerline{ 
\includegraphics[width=0.70\textwidth]{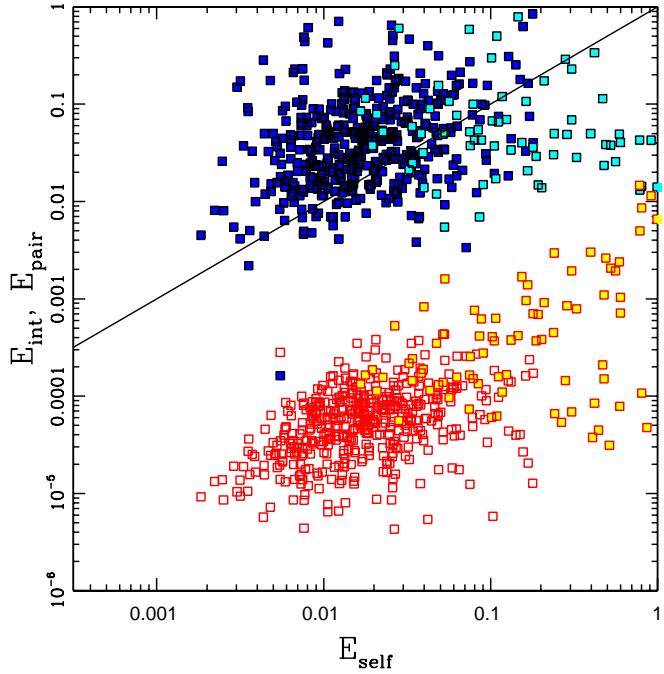} } 
\vskip-1.2truein
\caption{Energy scales in observed extrasolar multiple planet 
systems. The blue points show the energy available from pairwise
energy optimization for pairs of rocky planets in multiple systems,
plotted against the self-gravity of the bodies. The cyan points show
the analogous energy scale for pairs containing massive planets,
specifically those with $(m_1+m_2)>30M_\oplus$. For pairs of rocky
planets, the open red points show the energy of interaction between
planetary pairs, where $E_{\rm int}$ $\propto Gm_1m_2/(a\Delta)$. The
yellow points show the analogous energy scales for massive planets. }
\label{fig:enerplot} 
\end{figure} 

The underlying idea of pairwise equilibrium is that planetary systems
can attain lower energy states by optimizing their mass ratios, where
the critical point corresponds to nearly equal masses (specifically,
the mass fraction given by equation [\ref{mfract}]). The energy
available through this optimization, presumably approached during the
epoch of planet formation, is illustrated in Figure \ref{fig:energy}.
However, the analysis thus far has not included the self-gravity of
the planets themselves or the potential energy due to gravitational
interactions between planets. This section assesses the relative sizes
of these additional energy contributions and shows that they are 
sub-dominant for the observed multi-planet systems of interest. 

The energy available due to re-arrangement of planetary mass is given
by equation (\ref{pairenergy}), where this energy is scaled relative
to the orbital energy. The dimensionless self-gravity of the planets,
again scaled relative to their orbital energy, is given by 
\be
E_{\rm self} = \selfg {G \mplan^2 \over R_{\rm p}} 
\left( {2a_p \over GM_\ast \mplan} \right) = 
2\selfg\,{a_p\over R_{\rm p}}\,{\mplan \over M_\ast} \,, 
\label{selfenergy} 
\ee
where $\selfg$ is a dimensionless constant of order unity and depends
on the interior structure of the planet. Finally, the (dimensionless)
energy of interaction between planets in adjacent orbits can be
written in the form 
\be
E_{\rm int} = \beta {G m_1 m_2 \over a_2-a_1} 
\left[ {2 \langle a_p \rangle \over GM_\ast (m_1 +m_2)} \right] = 
\beta {\mu \over M_\ast} {a_1 + a_2 \over a_2-a_1} \lta 
{1 \over \Delta} \left({\mplan \over M_\ast}\right) \,,
\label{interenergy} 
\ee
where $\beta$ is another dimensionless factor of order unity 
(see \citealt{md} for further detail). 

The energy scales for adjacent planetary pairs residing in observed
multi-planet systems are depicted in Figure \ref{fig:enerplot}. 
All of the energies are scaled to the orbital energy and are thus
dimensionless. The upper collection of points shows the energy
available from pairwise equilibrium (using equation
[\ref{pairenergy}]) as a function of the summed self-gravity of the
two constituent planets (see equation [\ref{selfenergy}]). The dark
blue points show the results for pairs of rocky planets, whereas the
cyan points correspond to planets with higher masses, namely
$(m_1+m_2)>30M_\oplus$. The solid black curve marks the line of
equality. Notice that the majority of the blue points fall above the
solid curve, so that the pairwise energy dominates over that of
self-gravity. In contrast, more cyan points fall below the solid curve
than above it. This result suggests that the self-gravity of the
planets is relatively unimportant during the formation process for
rocky planets (blue points), but self-gravity plays a more important
role in the assembly of the largest planets (cyan points). Moreover,
the neglect of self-gravity represents a valid approximation for rocky
planets in multiple systems, whereas it should be included in systems
containing larger Jovian planets. The analysis of this paper is thus
limited to the former type of planetary system.

This treatment has ignored both the energy and angular momentum of the
planetary spins. Although the rotation rates of exoplanets are
generally not known, the rotational speeds are constrained to be less
than break up velocity. As a result, the rotational kinetic energy
must be smaller than that of self-gravity, which in turn is much less
than the pairwise equilibrium energy considered in this paper, as
shown in Figure \ref{fig:enerplot}. The neglect of spin in the energy
budget is thus justified. 

Figure \ref{fig:enerplot} also shows the interaction energy between
pairs of planets (from equation [\ref{interenergy}]) as a function of
self-gravity. The red points show the results for rocky planets,
whereas the yellow points correspond to the larger planets
$(m_1+m_2>30M_\oplus)$. The red points fall well below the line of
equality. Moreover, the red points (self-gravity) fall even farther
below the blue points (pairwise energy).  This latter result can be
interpreted as follows: The leading order expression for the pairwise
energy has the form $E_{\rm pair}\approx3\Delta^2/16$, and that for
the interaction energy becomes $E_{\rm int}\approx{\mplan}/(M_\ast\Delta)$. 
These two (dimensionless) energies are equal when the planetary
separation decreases to the value 
\be
\Delta = \left( {16\over3}{\mplan\over M_\ast} \right)^{1/3} = 
2^{4/3} {R_H \over a}\,.
\ee
In other words, the spacing of adjacent orbits must be as small as
$K=2^{4/3}$ for the interaction energy to dominate. Of course, given
that these systems experience a large number of dynamical time scales,
planetary interactions can affect system stability for wider spacing.
As discussed previously, observed systems have $K\approx10-20$, and
$K\gta10$ is required for long-term stability. The finding that the
pairwise energy is much greater than the interaction energy is thus
expected from considerations of orbital stability and from the
observed spacing of planetary orbits.

\subsection{Orbital Spacing} 
\label{sec:spacing} 

The discussion thus far has considered the orbital spacing $\Lambda$
to be fixed. In this section, we consider the observed distribution of
orbital spacings and place approximate constraints on the expected
range of values. One criterion is that the planets must not be too
close together in order to remain stable. In addition, if the planets
are spaced as tightly as possible subject to stability constraints ---
within a few Hill radii of each other --- the disk will not have
enough mass to form them at their final locations. As a result, we can
obtain a rough estimate for orbital spacing based on considerations of
the total mass budget in the disk. {Note that this order of
magnitude estimate does not include possibile redistribution of mass 
within the disk. Such rearrangement could take place by moving raw
material through the disk before the epoch of planet formation and/or
by changing the orbital elements of assembled planets through migration.}

\begin{figure} 
\centerline{ 
\includegraphics[width=0.70\textwidth]{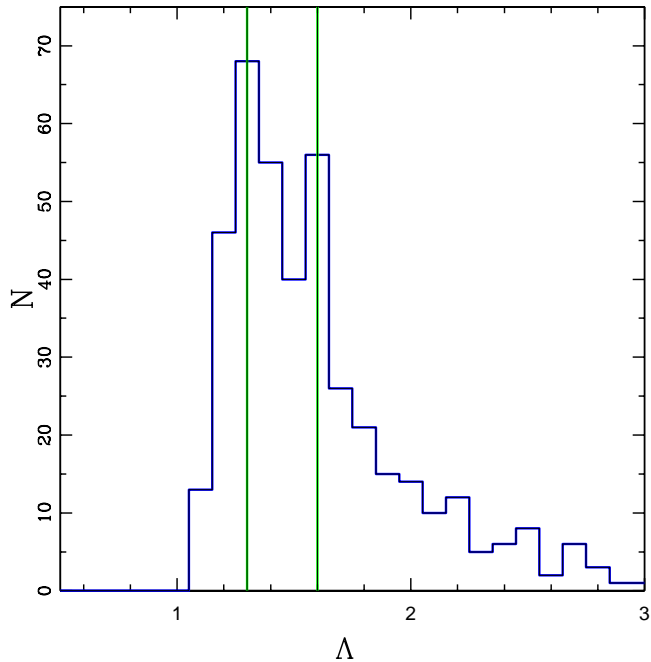} } 
\vskip-1.2truein
\caption{Distribution of orbital spacing parameter $\Lambda$ =
$a_2/a_1$ for observed pairs of planets. The blue histogram shows the
spacing parameters for the sample of observed multi-planet systems.
The green vertical lines delimit the range of $\Lambda$ parameters
expected for equally spaced planets constrained by the available mass
of rocky material in the disk (under the assumptions of equal mass
bodies and minimal planetary migration). } 
\label{fig:xlhist} 
\end{figure}  

The mass scale for rocky planets is observed to be of order
$\mplan\approx10M_\oplus$ \citep{zhu2018} and is remarkably consistent
within planetary systems. This mass scale can be derived from
considerations of pebble accretion taking place within the
circumstellar disks that gives birth to planets, as outlined in
Appendix \ref{sec:rockmass} (see also \citealt{lambrechts,bitsch} and
references therein). Since the total mass of the disk is finite, the
total number of rocky planets that can be produced is limited. The
rocky mass within a disk can be written in the form 
\be
M_{\rm rock} = Z M_{\rm disk} \lta \,Z\,{M_\ast\over10} \,, 
\ee
where $Z\sim0.01$ is dust to gas mass ratio and we have assumed that
the maximum disk mass is about 10\% of the stellar mass. This latter
fraction can be understood in terms of disk stability
\citep{starmodes} and is consistent with the upper limits on disk
masses in observed systems \citep{hartmann}. Under the assumption 
that all of the heavy elements in the disk can be utilized for planet
formation, the maximum number $\nump$ of rocky planets is given by  
\be
\nump = {Z M_\ast \over 10 \mplan} \approx 16 - 30 \,. 
\ee
Combining this result with the mass profile from equation 
(\ref{sigmar2}), we can solve for the spacing parameter to find 
\be
\ln \Lambda \approx {1 \over \nump} \ln(a_{\rm out}/a_{\rm in}) 
\qquad {\rm or} \qquad \Lambda \approx 
\left( {a_{\rm out} \over a_{\rm in}} \right)^{1/\nump} \,, 
\ee
where $a_{\rm out}$ and $a_{\rm in}$ represent the outer and inner
edges of the disk. For typical disk parameters $a_{\rm out}$ = 100 
AU and $a_{\rm in}$ = 0.05 AU, we find $\Lambda\approx1.3-1.6$. 

Figure \ref{fig:xlhist} shows the observed distribution of orbital
spacings for the planetary pairs in our comparison sample. For the
sake of definiteness, this compilation includes only pairs with total
mass $m_1+m_2\le30M_\oplus$, roughly corresponding to superearth
planets and thereby excluding Jovian planets.  The observed spacing
distribution (blue histogram) shows a peak near $\Lambda\sim1.4$. The
vertical green lines in the figure show the benchmark values $\Lambda$
= 1.3 and 1.6 derived above on the basis of the available mass in
rocky material. About 54\% of the pairs in the observational sample
fall in this interval ($1.3\le\Lambda\le1.6$), whereas about 77\% of
the pairs fall within the wider range $1.2\le\Lambda\le1.8$.  The
observed distribution of orbital spacings in these systems is roughly
consistent with that expected from considerations of the mass supply.
Nonetheless, the distribution displays a tail corresponding to wider
spacings, where $\sim$13\% of the pairs have $\Lambda>2$. Such large
values of $\Lambda$ could be explained by missing planets, planetary
migration, and/or the rearrangement of disk material before planets 
are formed (see, e.g., \citealt{hansen}). In any case, the full spacing 
distribution requires additional explanation and should be considered
in future work. 

\section{Conclusion} 
\label{sec:conclude} 

This paper formulates and solves a new type of Darwin problem that
determines the lowest energy (tidal equilibrium) states for
multi-planet systems subject to a collection of constraints. 
The optimized system architectures have properties that are 
{roughly consistent} with those of observed planetary systems. 
This section presents a summary of our specific results (Section
\ref{sec:summary}) and a discussion of several unresolved issues
(Section \ref{sec:discussion}).

\subsection{Summary of Results} 
\label{sec:summary} 

We first consider two planet systems subject to constraints of
constant angular momentum and fixed orbital spacing (Section
\ref{sec:double}). The minimum energy state of the system corresponds
to planets with nearly equal masses on circular orbits ($e_j=0$) in
the same plane ($i=0$).  The mass fraction $f\approx1/2$, so that
$m_1\approx{m_2}$, and varies slowly with orbital spacing as given by
equation (\ref{mfract}). These properties are realized in the observed
sample of extrasolar planetary systems found by the {\it Kepler}
mission.

For systems with more than two planets, where the total mass can be
freely distributed among the planets, no (global) minimum energy state
exists (Appendix \ref{sec:equilthree}). The system can evolve to a
lower energy state by consolidating the planets, thereby reducing
their number. Observed multiple planet systems do not reside in states
corresponding to global energy minima, so that some additional
consideration must act to define their properties.

For multiple planet systems, we introduce the concept of pairwise
equilibrium, where each pair of planets in adjacent orbits attains the
minimum energy state appropriate for the pair (Section
\ref{sec:multiple}), subject to conservation of angular momentum and
fixed orbital spacing. With each pair in this type of equilibrium, the
resulting planetary system has properties much like those observed in
the {\it Kepler} sample: circular orbits confined to a single plane,
with regular orbital spacing and nearly uniform planetary masses. To
higher order, the planetary mass is a slowly increasing function of
semimajor axis (see equations [\ref{etadef}] and [\ref{etamass}]). 

{The optimal configurations found in this paper can be compared
to observed planetary systems (Section \ref{sec:observe}).} The
pairwise equilibrium states have nearly equal mass planets on
regularly spaced orbits. This type of system has an effective surface
density of the form $\sigma\propto$ $r^{-2}$ and mass profile
$M(r)\propto\ln(r)$, which differs substantially from the usual form
$\sigma\propto{r^{-3/2}}$ of the MMSN or the form
$\sigma\propto{r^{-1}}$ inferred for circumstellar disks.  
{The rocky component of observed multi-planet systems have 
similar surface density and mass profiles for orbital periods 
$P<100$ days (see Figure \ref{fig:massprofile}). These profiles 
arise from nearly equal mass planets on regularly spaced orbits, 
a configuration that has been suggested in previous studies 
\citep{rowe2014,puwu,millholland,songhu,weiss1,weiss2}.} 

In the multiple-planet systems of interest, the typical mass
$\mplan\sim10M_\oplus$ can be understood as the isolation mass
produced through pebble accretion (Appendix \ref{sec:rockmass}).
Given this planetary mass scale, the energy difference that can be
realized by systems approaching a tidal equilibrium state is larger
than both the self-gravity of the individual bodies and the
interaction energy between planets (see Figure \ref{fig:enerplot}). 
As a result, during the process of planet formation, considerations of
pairwise equilibrium are energetically important. For Jovian planets,
however, the self-gravity plays a more dominant role. This apparent
dichotomy is consistent with the previous finding that intra-system
uniformity of planetary masses breaks down in the presence of giant
planets \citep{songhu}.

\subsection{Discussion}
\label{sec:discussion} 

This paper has shown that the minimum energy configurations available
to multi-planet systems require nearly equal planetary masses, along
with circular orbits in the same plane.  Although this finding is
consistent with the observed properties of planetary systems, it
provides only a partial explanation, and a number of issues remain
unresolved: 

The optimization procedure of this work shows that the critical state
for two planet systems has the mass fraction $f\approx1/2$ given by
equation (\ref{mfract}). The determination of planetary masses takes
place (by definition) during the epoch of planet formation. However,
the existence and properties of the minimum energy state do not
explain how the process of planet formation realizes this goal. On a
related note, the assumption of constant mass for planetary pairs is
an idealization.  Moreover, in planetary systems containing giant
planets, the self-gravity of the planets dominates the pairwise
effects considered in this paper, and observations show that such
systems no longer display the same regular features (e.g.,
\citealt{songhu}). As a result, some process --- yet to be determined
--- drives some forming planetary systems to produce giant planets
while other systems produce multiple superearths.

Another unresolved issue concerns the orbital spacing of the planets.
This work shows that planetary systems prefer nearly equal masses for
fixed orbital spacing, but does not provide an explanation for the
size of the observed spacing. Previous work shows that orbits cannot
be too close without rendering the systems dynamically unstable (e.g.,
\citealt{gladman,puwu}), thereby providing a lower limit on orbital
spacing. Considerations of statistical mechanics can produce
consistent spacing distributions \citep{tremaine15}, but require
particular choices for the relevant parameters (see also
\citealt{mogavero,pakter}). Using limits on the available disk mass
and radius, this paper constructs estimates for the orbital spacing
parameters that are roughly consistent with those observed (Section
\ref{sec:spacing}). The spacing could be tighter than these estimates,
however, if some migration process moves planets together after they
are formed, or transports rocky material inward before it becomes
incorporated into planets. This work also indicates that equal mass
planets on more widely separated orbits are energetically preferred.
Nonetheless, we would like to know the details concerning how the
process of planet formation produces the observed distribution of
orbital spacing (see Figure \ref{fig:xlhist}).

Planets with equal masses and regular spacing, as found in both
observed systems and the minimum energy configurations of this paper,
correspond to an effective surface density profile $\sigma(r)\sim$
$r^{-2}$, which is steeper than that of both observed circumstellar
disks and the MMSN. As a result, some process must redistribute the
rocky mass within these systems, either as raw material before
(perhaps during) planet formation or later as planetary bodies with
their final masses. The timing and mechanism(s) of this redistribution
process also remains unresolved. In addition, the observed mass
distribution shows a power-law tail inside the nominal inner radius of
the disk (see Figure \ref{fig:massprofile}). The manner in which this 
region is populated by planets poses another interesting problem. 

The model of pairwise equilibrium developed here applies to
compact planetary systems, such as those observed by {\it Kepler}.
However, we currently have limited sensitivity to additional planets
on wider orbits. If these planetary systems can reach the
energy-optimized states considered here, then the masses of additional
exterior planets would slowly increase with semimajor axis (for fixed
orbital spacing; see Section \ref{sec:multiple}; see also
\citealt{jiang}). Reaching such equilibrium states requires the planet
formation process to communicate across the entire planetary
system. Future developments in planet formation theory and extended
observations of planetary systems are necessary to determine if this
is the case. On a related note, the observed systems could have inner
planets that are closely spaced, with a wide gap separating a
collection of additional planets. In this context, the two (separated)
sets of planets could reach separate states of pairwise
equilibrium. If the outer planets are sufficiently massive, however,
then their self-gravity would dominate over the pairwise energy, and
the systems are less likely to be well-ordered. 

Finally, we note that the Galilean satellites also display nearly
uniform masses, regular orbital spacing, low eccentricities, and
coplanar orbits aligned with the equatorial plane of Jupiter. This
remarkable regularity suggests that these moons represent a miniature
solar system, whose formation requires an explanation (e.g.,
\citealt{lunine}).  The pairwise equilibrium solutions of this paper
lead to similar configurations and could thus operate on smaller mass
and size scales. 

The open issues outlined above, along with the diversity of the
planetary systems that are ultimately produced, highlight the
complicated nature of the planet formation process. In the face of
such complexity, one way forward is to identify some type of
organizing principle that allows us to understand and/or constrain the
possible outcomes. The pairwise minimum energy configurations found in
this paper provide one avenue toward this end.

\medskip 
\textbf{Acknowledgments:} 
We would like to thank Konstantin Batygin, Juliette Becker, Tony
Bloch, Greg Laughlin, Erik Petigura, Darryl Seligman, and Chris
Spalding for useful discussions. We also thank an anonymous referee
for constructive comments. This work was supported by the University
of Michigan.

\appendix
\section{Absence of Equilibrium States for Three Planet Systems} 
\label{sec:equilthree} 

This Appendix considers the possible equilibrium state for a three
planet system in which the mass can be distributed among the three
planets and no pairwise constraints are applied. In this case, no
equilibrium state for the three planet system exists. Equivalently,
the equilibrium state requires the middle planet to have the same
orbit as the inner planet, thereby reducing the system to two bodies.

For three planets, the mass fractions can be written in the form 
\be
f = {m_1 \over \mtot}\,, \qquad 
g = {m_2 \over \mtot}\,, \qquad 
1-f-g = {m_3 \over \mtot}\,, 
\ee
where $\mtot$ is the total mass of the three planets.  Here we
specialize to the case of circular orbits, all confined to the same
orbital plane. The semimajor axes of the orbits are then given by 
\be
a_1 = a \,, \qquad a_2 = \Lambda a \,, \qquad {\rm and} 
\qquad a_3 = \Lambda \Gamma a \,,
\ee
where the factors $\Lambda$ and $\Gamma$ are considered fixed. For
this reduced problem, the dimensionless energy and angular momentum
take the form 
\be
E = - {1 \over a} \left[ f + {g \over \Lambda} + 
{1-f-g \over \Lambda \Gamma} \right] \,,
\ee
and
\be
L = \sqrt{a} \left[ f + g \sqrt{\Lambda} + 
(1-f-g) \sqrt{\Lambda \Gamma} \right] \,.
\ee
To minimize the energy subject to conservation of angular momentum, 
we must find the critical point of the function $F = E + \lambda L$, 
where $\lambda$ is the Lagrange multiplier. The derivatives for the 
mass fractions have the form 
\be
{\partial F \over \partial f} = 
- {1 \over a} \left[ 1 - {1 \over \Lambda \Gamma} \right] 
+ \lambda \sqrt{a} \left[ 1 - \sqrt{\Lambda \Gamma} \right] = 0
\ee
and 
\be
{\partial F \over \partial g} = - {1 \over a} 
\left[ {1\over\Lambda} - {1\over\Lambda\Gamma} \right] + 
\lambda \sqrt{a} \left[ \sqrt{\Lambda} - \sqrt{\Lambda\Gamma} \right] 
=0\,. 
\ee
Solving both equations for the quantity $\lambda a^{3/2}$, we find 
\be
- \lambda a^{3/2} = 
\left[ 1 - {1 \over \Lambda \Gamma} \right] 
\left[ \sqrt{\Lambda \Gamma} - 1\right]^{-1} = 
\left[ {1\over\Lambda} - {1\over\Lambda\Gamma} \right] 
\left[ \sqrt{\Lambda\Gamma} - \sqrt{\Lambda} \right]^{-1} \,. 
\ee
The second equality determines the required relation between 
the separation factors $\Lambda$ and $\Gamma$, and can be 
simplified to the form 
\be
\Lambda \sqrt{\Gamma} + \sqrt{\Lambda} = \sqrt{\Gamma} + 1 \,. 
\ee
This expression implies that $\Lambda=1$. In other words, any solution
for the lowest energy state of the three planet system requires the
middle planet to have the same orbit as the inner planet, thereby
producing a two planet system. For three separate planets, which 
requires $\Lambda>1$, no tidal equilibrium state exists. 

\begin{figure} 
\centerline{ 
\includegraphics[width=0.70\textwidth,trim=3cm 4cm 3cm 4cm,clip=true]
{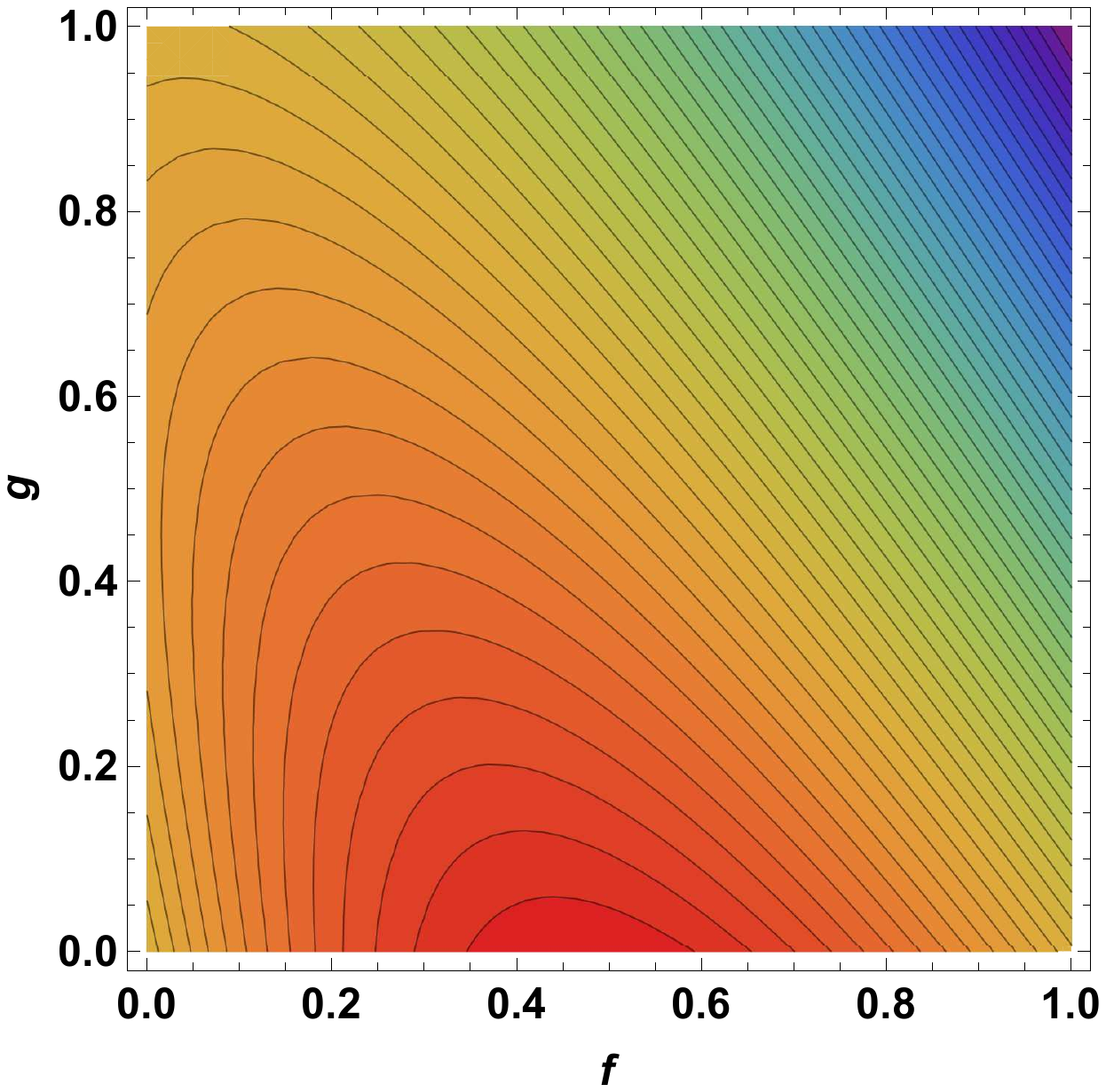}}  
\vskip-72pt 
\caption{Contour plot showing the energy for a three planet system
as a function of the mass fractions $f$ and $g$ of the inner and
middle planet. The orbital spacings are fixed to be $\Lambda$ =
$\Gamma$ = 5/4.  The minimum energy state corresponds to the limit
$g\to0$, where the middle planet has zero mass and the mass fraction
$f$ of the inner planet is that of the two planet system (see equation
[\ref{mfract}]). }
\label{fig:fgcontour} 
\end{figure}  

This trend is illustrated in Figure \ref{fig:fgcontour}, which shows
the iso-energy contours for a three planet system as a function of the
mass fractions $f$ and $g$ of the inner two planets (the mass fraction
of the third planet is determined by conservation of mass to be
$[1-f-g]$). The mass fraction $g$ of the middle planet approaches the
$f$-axis for the minimum energy state, i.e., the mass of the middle
planet vanishes. In this $g\to0$ limit, the mass fraction $f$ of the
inner planet approaches the value appropriate for the two planet
system (from Section \ref{sec:double} and equation [\ref{mfract}]).
Note that the mass fraction of the outer planet is given by $1-f-g$, 
so that the physically relevant part of the diagram is confined to 
the region defined by $f+g\le1$. 

\section{Characteristic Mass Scale for Rocky Planet Formation} 
\label{sec:rockmass} 

This Appendix derives a characteristic mass scale for planet formation
based on the idea that planets accumulate mass through pebble
accretion (e.g., \citealt{lambrechts,bitsch}) which eventually shuts
down when the pressure gradient produced by the planet becomes
sufficiently large. As shown below, this benchmark mass scale falls in
the range $\mplan=4-8M_\oplus$, consistent with planetary masses
typically found in observed multi-planet systems.

The concept behind the isolation mass for pebble accretion is that the
mass of the growing planet (or planetary core) becomes large enough to
affect the pressure of the surrounding gas, and thereby provides a
pressure gradient that inhibits further growth. In order of magnitude, 
application of the Bernoulli equation indicates that the planet will 
influence the flow of gas parcels by increasing their speed, and this 
increase will become significant when 
\be
(\Delta v)^2 \sim {2G \mplan \over b} \sim \sound^2 \,,
\label{balance} 
\ee
where $\mplan$ is the mass of the growing planet (core), 
$b$ is the impact parameter of the fluid parcel trajectory, 
and $\sound$ is the sound speed in the gas. 

The planet receives pebbles from within its Hill sphere, so we take
the impact parameter to be comparable to the Hill radius.  This choice
allows us to write the stopping condition in the form
\be
2G \mplan = \pfact\,\sound^2 b 
= \pfact\,\sound^2 \left({\mplan \over 3 M_\ast}\right)^{1/3} r
= \pfact\,\sound^2 \left({G \mplan \over 3 \Omega_{\rm k}^2 }\right)^{1/3} \,,
\ee
where $\Omega_{\rm k}$ is the Keplerian rotation rate of the disk and where
$\pfact$ is a dimensionless factor of order unity.  Solving for the
planetary mass yields 
\be
\mplan = {\pfact^{3/2}\over\sqrt{3}} \, {\sound^3 \over G \Omega_{\rm k}} \,.
\ee
Next we want to write the isolation mass in terms of the scale 
height $H$ of the disk. The standard definition has the form 
\be
{H \over r} = {\sound \over \Omega_{\rm k} r} \,, 
\ee
so that 
\be
\mplan = {\pfact^{3/2}\over\sqrt{24}} \, \left({H\over r}\right)^3 
\, {\Omega_{\rm k}^2 r^3 \over G} = 
{\pfact^{3/2}\over\sqrt{24}} \, \left({H\over r}\right)^3 M_\ast \,.
\label{mscalezero} 
\ee
Using the typical value $H/r\sim0.05$, we can write this 
expression in the form 
\be
\mplan = 8.5 M_\oplus \pfact^{3/2} \left({H/r\over 0.05}\right)^3 
\left({M_\ast\over1M_\odot}\right) \,. 
\label{mscale} 
\ee
For the observational sample of planets, the typical stellar mass 
$M_\ast\approx0.5-1 M_\odot$, so that the benchmark planetary mass 
is about $\mplan\approx4-8M_\oplus$. This value is roughly 
consistent with the mass scale of planets observed in multiple 
systems \citep{zhu2018}.

\label{lastpage}


\begin{thebibliography}{99}

\bibitem[Abramowitz \& Stegun(1972)]{abrasteg} 
Abramowitz, M., \& Stegun, I. A. 1972, Handbook of Mathematical
Functions (New York: Dover)

\bibitem[Adams \& Bloch(2015)]{ab2015}
Adams, F. C., \& Bloch, A. M. 2015, MNRAS, 446, 3676

\bibitem[Adams \& Bloch(2016)]{ab2016}
Adams, F. C., \& Bloch, A. M. 2016, MNRAS, 462, 2527

\bibitem[Alessi et al.(2017)]{alessi} 
Alessi, M., Pudritz, R. E., \& Cridland, A. J. 2017, MNRAS, 464, 428 

\bibitem[Andrews et al.(2009)]{andrews} 
Andrews, S. M., Wilner, D. J., Hughes, A. M., Qi, C., \& 
Dullemond, C. P. 2009, ApJ, 700, 1502

\bibitem[Batalha et al.(2011)]{batalha} 
Batalha, N. M., Borucki, W. J., Bryson, S. T., et al. 2011, ApJ, 729, 27

\bibitem[Becker \& Adams(2016)]{beckeradams16} 
Becker, J. C., \& Adams, F. C. 2016, MNRAS, 455, 2980 

\bibitem[Becker \& Adams(2017)]{beckeradams17} 
Becker, J. C., \& Adams, F. C. 2017, MNRAS, 468, 549

\bibitem[Bitsch et al.(2015)]{bitsch} 
Bitsch, B., Lambrechts, M., \& Johansen, A. 2015, A\&A, 582, 112

\bibitem[Borucki et al.(2010)]{borucki}
Borucki, W. J., Koch, D., Basri, G., et al. 2010, Sci, 327, 977

\bibitem[Chambers et al.(1996)]{chambers} 
Chambers, J. E, Wetherill, G. W., \& Boss, A. P. 1996, Icarus, 119, 261 

\bibitem[Chiang \& Laughlin(2013)]{chiang} 
Chiang, E., \& Laughlin, G. 2013, MNRAS, 431, 3444 

\bibitem[Counselman(1973)]{counselman} 
Counselman, C. C. 1973, ApJ, 180, 307 

\bibitem[Darwin(1879)]{darwin1} 
Darwin, G. H. 1879, The Observatory, 3, 79

\bibitem[Darwin(1880)]{darwin2} 
Darwin, G. H. 1880, Phil. Trans. R. Soc. A, 171, 713 

\bibitem[Desch(2007)]{desch} 
Desch, S. J. 2007, ApJ, 671, 878

\bibitem[Fabrycky et al.(2014)]{fabrycky}
Fabrycky, D. C., Lissauer, J. J., Ragozzine, D., et al. 2014, ApJ, 790, 146 

\bibitem[Fang \& Margot(2012)]{fangmargot} 
Fang, J., \& Margot, J.-L. 2012, ApJ, 761, 92

\bibitem[Gladman(1993)]{gladman} 
Gladman, B. 1993, Icarus, 106, 247  

\bibitem[Hadden \& Lithwick(2017)]{hadden} 
Hadden, S., \& Lithwick, Y. 2017, AJ, 154, 5 

\bibitem[Hansen \& Murray(2013)]{hansen}
Hansen, B.M.S., \& Murray, N. 2013, ApJ, 775, 53,

\bibitem[Hartmann(2007)]{hartmann} 
Hartmann L. 2007, Physica Scripta, 130, 014012 

\bibitem[Hayashi(1981)]{hayashi} 
Hayashi, C. 1981, Prog. Theor. Phys. Suppl., 70, 35 

\bibitem[Hesse(1872)]{hesse} 
Hesse, L. O. 1872, Die Determinanten elementar behandelt (Leipzig)

\bibitem[Hut(1980)]{hut1980}
Hut, P. 1980, A\&A, 92, 167

\bibitem[Hut(1981)]{hut1981}
Hut, P. 1981, A\&A, 99, 126 

\bibitem[Ida \& Lin(2010)]{idalin} 
Ida, S., \& Lin, D.N.C. 2010, ApJ, 719, 810

\bibitem[Jiang et al.(2015)]{jiang}
Jiang, I.-G., Yeh, L.-C., \& Hung, W.-L. 2015, MNRAS, 449, 65 

\bibitem[Kipping(2018)]{kipping} 
Kipping, D. 2018, MNRAS, 473, 784 

\bibitem[Lambrechts et al.(2014)]{lambrechts}
Lambrechts, M., Johansen, A., \& Morbidelli, A. 2014, A\&A, 572, 35

\bibitem[Laughlin et al.(2004)]{lba}
Laughlin, G., Bodenheimer, P., \& Adams, F. C. 2004, ApJ, 612, L73 

\bibitem[Levrard et al.(2009)]{levrard} 
Levrard, B., Winisdoerffer, C., \& Chabrier, G. 2009, ApJ, 692, 9

\bibitem[Lunine \& Stevenson(1982)]{lunine} 
Lunine, J. I., \& Stevenson, D. J. 1982, Icarus, 52, 14 

\bibitem[Malhotra(2015)]{malhotra} 
Malhotra, R. 2015, ApJ, 808, 71

\bibitem[Millholland et al.(2017)]{millholland} 
Millholland, S., Wang, S., \& Laughlin, G. 2017, ApJ Letters, 849, L33

\bibitem[Mogavero(2017)]{mogavero}
Mogavero, F. 2017, A\&A, 606, 79

\bibitem[Mordasini et al.(2009)]{mordasini}
Mordasini, C., Alibert, Y., \& Benz, W. 2009, A\&A, 501, 1139 

\bibitem[Mordasini(2018)]{mordasini18}
Mordasini, C. 2018, in Handbook of Exoplanets, Planetary Population
Synthesis, Springer, New York City, NY, USA, p. 143

\bibitem[Murray \& Dermott(1999)]{md}
Murray, C. D., \& Dermott, S. F. 1999, Solar System Dynamics 
(Cambridge: Cambridge Univ. Press) 

\bibitem[Obertas et al.(2017)]{obertas}
Obertas, A., Van Laerhoven, C., \& Tamayo, D. 2017,
Icarus, 293, 52

\bibitem[Pakter \& Levin(2018)]{pakter}
Pakter, R., \& Levin, Y. 2018, Phys. Rev. E, 97, 2221 

\bibitem[P{\'e}rez et al.(2012)]{perez}
P{\'e}rez, L. M., Carpenter, J. M., Chandler, C., et al. 2012, 
ApJ, 760, L17

\bibitem[Pu \& Wu(2015)]{puwu} 
Pu, B., \& Wu, Y. 2015, ApJ, 807, 44

\bibitem[Rowe et al.(2014)]{rowe2014} 
Rowe, J. F., Bryson, S. T., Marcy, G. W., et al. 2014, ApJ, 784, 45

\bibitem[Shu et al.(1990)]{starmodes} 
Shu, F. H., Tremaine, S., Adams, F. C., \& Ruden, S. P. 1990, 
ApJ, 358, 495 

\bibitem[Tremaine(2015)]{tremaine15} 
Tremaine, S. 2015, ApJ, 807, 157 

\bibitem[Tremaine \& Dong(2012)]{tredong}
Tremaine, S., \& Dong, S. 2012, AJ, 143, 94 

\bibitem[Van Eylen \& Albrecht(2015)]{vaneylen}
Van Eylen, V., \& Albrecht, S. 2015, ApJ, 808, 126

\bibitem[Wang(2017)]{songhu}
Wang, S. 2017, RNAAS, 1, 26 

\bibitem[Weidenschilling(1977)]{weiden}
Weidenschilling, S. J. 1977, MNRAS, 180, 57 

\bibitem[Weiss et al.(2018a)]{weiss1} 
Weiss, L. M., Marcy, G. W., Petigura, E. A., et al. 2018a, AJ, 155, 48

\bibitem[Weiss et al.(2018b)]{weiss2} 
Weiss, L. M., Marcy, G. W., Petigura, E. A., et al. 2018b, AJ, 156, 254

\bibitem[Winn \& Fabrycky(2015)]{winnfab}
Winn, J. N., \& Fabrycky, D. C. 2015, ARA\&A, 53, 409

\bibitem[Wolfgang et al.(2016)]{wolfgang} 
Wolfgang, A., Rogers, L. A., \& Ford, E. B. 2016, ApJ, 825, 19 

\bibitem[Wu et al.(2019)]{wu2019} 
Wu, D.-H., Zhang, R. C., Zhou, J.-L., \& Steffen, J. H. 2019, 
MNRAS 484, 1538–1548 (2019)

\bibitem[Zhu et al.(2018)]{zhu2018} 
Zhu, W., Petrovich, C., Wu, Y., Dong., S., \& Xie, J. 2018, 
ApJ, 860, 101 

\end{thebibliography}
\end{document}